\documentclass[prb,aps,amsmath,amssymb,showpacs,1superscriptaddress,floatfix,twocolumn]{revtex4}
\usepackage{graphicx,epsfig}
\newcommand{\bbf}{}
\newcommand{\gbf}{}

\begin{document}
\title{Failure of  mean-field approach in out-of-equilibrium Anderson model} 
\author{B. Horv\'ath$^1$, B. Lazarovits$^{1,2}$, O. Sauret$^1$, G. Zar\'and$^1$}
\affiliation{
$^1$Theoretical Physics Department, Institute of Physics, Budapest University of 
Technology and Economics, Budafoki \'ut 8, H-1521 Hungary\\
$^2$Research Institute for Solid State Physics and Optics of the Hungarian Academy 
of Sciences,
Konkoly-Thege M. \'ut 29-33., H-1121 Budapest, Hungary}

\begin{abstract}
To explore the limitations of the mean field approximation, 
frequently used in \textit{ab initio} {\bbf molecular electronics} 
calculations, we study an out-of-equilibrium 
Anderson impurity model in a scattering formalism. 
We find regions in the parameter space where both
magnetic and non-magnetic solutions are stable.
We also observe a hysteresis in the non-equilibrium 
magnetization and current as a function of the applied bias voltage. 
The mean field method also predicts incorrectly
local moment formation for large biases and a spin
polarized current, and unphysical kinks
appear in various physical quantities. The mean field
approximation thus fails {\bbf in every region where} it predicts
local moment formation.
\end{abstract}


\maketitle


The Anderson impurity model\cite{anderson} 
(AIM) has been the subject of great theoretical and experimental 
interest in the past decades (for a review see Ref.~\onlinecite{hewson}). 
There is a number of experimental systems including 
quantum dots, or single atoms and molecules contacted by 
leads, which  provide experimental realizations of various versions of the 
AIM under out-of-equilibrium conditions.  These systems are not 
just prototypes of out of equilibrium systems but 
a theoretical understanding of them would be  crucial 
for future molecular electronics and mesoscopic applications.

Anderson constructed his famous model in Ref.~\onlinecite{anderson} to 
describe local moment formation  and solved it within the
mean-field (MF) approximation. Within this approximation, 
he found a phase transition to a state where magnetic
moments are formed. Further work revealed that, in reality, 
quantum fluctuations of this 
local moment lead to the formation of 
a Kondo-singlet between the impurity and conduction 
electrons\cite{hewson} at low temperature, where the impurity spin is thus
completely screened. The spontaneous symmetry breaking found by Anderson 
is thus  an artifact of the mean field approximation.
Nevertheless,  the MF treatment indicates clearly the regions of strong 
correlations,  and it can also serve as a starting point for 
accurate approximations as  in the  local 
moment approach\cite{log1} (LMA) or 
interpolative perturbation  theory\cite{martin,ali1} (IPT).  The latter
approach  can easily
be generalized to non-equilibrium situations\cite{yeyati,ali2} 
using Keldysh formalism\cite{rammer}.  

In lack of more accurate methods, the mean field approximation is also 
used in molecular electronics calculations,
where LDA or eventually Hartree-Fock equations are solved in a scattering
state or Keldysh approach to describe moment 
formation\cite{nano}. However, it is not clear at all, how reliable these approximations are. 
The purpose of this paper is to shed some light on  the weaknesses of the non-equilibrium mean field 
approach on the simplest possible  test case, the out of equilibrium Anderson 
model, and to show  where usual ab initio calculations should fail.
Our conclusion is that the mean field approach fails qualitatively and
quantitatively essentially everywhere where it predicts local moment formation.
Our study, which is based on the scattering state formalism, is 
complementary to the recent work of Komnik and Gogolin\cite{komnik}, 
who used a Green's function formalism to study the mean field equations of the
non-equilibrium Anderson model. As we shall see, in the strongly
correlated regions several artifacts emerge such as non-equilibrium driven
spontaneous symmetry breaking as well as multiple stable solutions 
which lead to the appearance of hysteresis. These instabilities are 
probably also parts of the reasons, why non-equilibrium IPT suffers from all
kinds of instabilities. These instabilities were avoided in previous 
works by applying spin-independent approximations\cite{horvatic} 
or using an interpolative self-energy\cite{martin} or both\cite{kotliar}.

The non-equilibrium AIM Hamiltonian 
consists of four parts. 
The first part describes a single impurity level with energy 
$\varepsilon_{d}$
and an on-site Coulomb interaction ($U$)
\begin{equation}
   H_{d}=\sum_{\sigma=\uparrow,\downarrow}
         \varepsilon_{d} d_{\sigma}^{\dagger} d_{\sigma} 
         + Un_{\uparrow} n_{\downarrow}\;,\label{EqDot}	
\end{equation}
where $d_{\sigma}^{\dagger}$ and $d_{\sigma}$ are the creation and 
annihilation operators of the impurity electrons corresponding to 
spin state \(\sigma\) and \( n_{\sigma}=d_{\sigma}^{\dagger} d_{\sigma}\). 
The second and third terms 
describe the left (L) and right
(R) leads {which we model} by tight-binding chains,
\begin{equation}
  H_{\alpha}=\sum_{k,\sigma}
               \left(
                  -2\tilde{t}\cos k + \mu_{\alpha}
               \right)
               c_{k\alpha\sigma}^{\dagger} c_{k\alpha\sigma}\;.\label{EqLead}
\end{equation}
Here $\alpha\in(L,R)$, $c_{k\alpha\sigma}^{\dagger}$ and $c_{k\alpha\sigma}$
are the creation and annihilation 
operators of a conduction electron of wave number
$k\in \{0,\pi\}$ in lead $\alpha$, $\tilde{t}$ is the hopping
along the leads, and $\mu_{\alpha}$ is the
chemical potential of the left or the right lead.
The chemical potentials of the two leads are different due to
a finite bias voltage leading to a non-equilibrium situation.
The fourth part, \(H_{t}\), describes the tunneling
between  the leads and the impurity 
\begin{equation}
	H_{t}=V\sum_{\sigma}
               \big[
                 d_{\sigma}^{\dagger}(V_{-}c_{-1\sigma}+V_{+}c_{1\sigma})+ h. c.]\;.\label{EqTun}
\end{equation}
Here $V_{\mp}$ are the hybridization matrix elements between the impurity and the {\bbf left and the right}
leads, and $c_{l=\pm1\sigma}$ 
denote the conduction electron annihilation 
operators on the sites next to the impurity; sites along the left and right chains
are labelled by $l=\{-\infty,\dots,-1\}$ and $l=\{1,\dots,\infty\}$, respectively.
{\bbf For the sake of simplicity, here we study a symmetrical situation, 
$V_{-}=V_{+}=V$, but our conclusions are rather independent of this assumption.} 

To study the Hamiltonian above we used a mean-field approximation
{\bbf and replaced} the impurity term as
\begin{equation}
	H_{d}\rightarrow H_{d}^{MF}=
        \sum\limits_{\sigma}\left(\varepsilon_{d}+U\langle 
        n_{-\sigma}\rangle\right)d_{\sigma}^{\dagger}d_{\sigma}\;.
       \label{EqMF}
\end{equation}
The non-equilibrium expectation value of the occupation
numbers, $\langle n_{\sigma}\rangle$, 
in Eq.~(\ref{EqMF}) can be obtained by solving 
self-consistent equations discussed later.
The expression $\varepsilon_{d}+U\langle n_{-\sigma}\rangle$ can be 
regarded as an effective energy level of spin-$\sigma$ electron.
The hybridization between impurity and conduction electrons gives rise to a
finite lifetime for impurity states, reflected in the broadening
of the effective impurity levels with a
finite width, \(\Gamma=2\pi V^{2}\rho_{0}=V^{2}/\tilde{t}\) 
where $\rho_{0}$ is the density of states (DOS) of the conduction electrons at the Fermi-level 
{\bbf of the half-filled leads.}

To evaluate the non-equilibrium expectation values, 
$\langle n_{\sigma}\rangle$, we shall use a scattering formalism.
The annihilation operator of the left-coming scattering state
of energy $\varepsilon$ and spin $\sigma$ can be expressed as
\begin{eqnarray}\label{wvfunc}
c_{\sigma} 
(\varepsilon)
& = &
 \sum_{l<0}c_{l\sigma}e^{ikR_l}+\alpha_{\sigma}(\varepsilon)\sum_{l<0}c_{l\sigma}e^{-ikR_l}+
{}\nonumber\\
 & & {}+\beta_{\sigma}(\varepsilon)\sum_{l>0}c_{l\sigma}e^{ik'R_l}+\gamma_{\sigma}(\varepsilon) d_{\sigma}\;.
\end{eqnarray}
Here $c_{l\sigma}$ is the annihilation operator of the $l$th site
and $\alpha_{\sigma}(\varepsilon)$, $\beta_{\sigma}(\varepsilon)$ 
and $\gamma_{\sigma}(\varepsilon)$ 
are the reflection, transmission and dot coefficients, respectively, 
which we obtain by solving the corresponding Schr\"odinger equation.
Since the energy is conserved in course of the scattering process, the wave numbers 
$k$ and $k'$ are connected by the dispersion relation
$\varepsilon=-2\tilde{t}\cos k +\mu_{L}=-2\tilde{t}\cos k' + \mu_{R}$.
Waves coming from the right hand side can be constructed in a similar way. 
The occupation numbers are then calculated from the dot coefficients, 
$\gamma_{\sigma}(\varepsilon)$ and $\gamma'_{\sigma}(\varepsilon)$ 
corresponding to the states coming from the left and the right, respectively,
\begin{equation}
\langle n_{\sigma}\rangle = \rho_{0}\int_{-\infty}^{\infty}\mathrm{d}\varepsilon\left(|\gamma_{\sigma}(\varepsilon)
|^{2}f_{L}(\varepsilon)+|\gamma'_
{\sigma}(\varepsilon)|^{2}f_{R}(\varepsilon)\right)\;,\label{scint}
\end{equation}
where the DOS of leads is approximated by 
its value at Fermi-level, and  
$f_{\alpha}(\varepsilon)\equiv f(\varepsilon-\mu_{\alpha})$ 
is the Fermi function.

In the large bandwidth approximation ($\tilde{t}\rightarrow\infty$, 
$\Gamma=V^{2}/\tilde{t}$ finite) the occupation numbers become
\begin{equation}
\langle n_{\sigma}\rangle=\frac{\Gamma}{2\pi}\int\limits_{-\infty}^{\infty}\frac{f_{L}(\varepsilon)+f_{R}(\varepsilon)}{(\varepsilon-\varepsilon_{d}-U\langle n_{-\sigma}\rangle)^{2}+\Gamma^{2}}\mathrm{d}\varepsilon\label{EqMain}\;,
\end{equation}
which simplifies further for zero temperature as 
\begin{equation}
\langle n_{\sigma}\rangle=
\sum\limits_{\alpha\in(L,R)}\frac{1}{2\pi}\cot^{-1}
\bigg(\frac{\varepsilon_{d}+
U\langle n_{-\sigma}\rangle-\mu_{\alpha}}
{\Gamma}\bigg)\;.\label{EqMain0}
\end{equation}
{\bbf These self-consistent equations can also be obtained using the Keldysh 
formalism\cite{komnik}.}
Eqs.~(\ref{EqMain}) or (\ref{EqMain0}) are solved 
iteratively by assuming that the applied bias is symmetrical 
$\mu_{L}=\mu/2$ and $\mu_{R}=-\mu/2$.
In these calculations the bias voltage is increased or
decreased gradually and the local stability of the solutions
is always checked.


First, let us discuss the  bias-dependence of the occupation numbers 
at $T=0$.
\begin{figure}
\includegraphics[width=200pt]{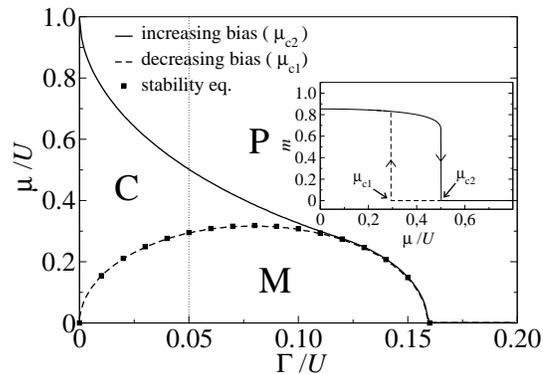}
\caption{Magnetic (M), coexistence (C) and paramagnetic (P) regions
as a function of $\mu/U$ and $\Gamma/U$ values for $\varepsilon_{d}/U=-1/2$.
The boundaries of the regions obtained by increasing and decreasing
bias are plotted with dashed and dotted lines, respectively.
The exactly evaluated boundary curve is indicated by 
squares. Inset: Magnetization as a function of increasing and 
decreasing bias voltage for $\Gamma/U=0.05$ (along the dotted line in the main
figure)
}\label{FigMuGamma}
\end{figure}
In the inset of Fig.~\ref{FigMuGamma} the magnetization 
$m=\langle n_{\uparrow}\rangle-\langle n_{\downarrow}\rangle$ is plotted
both for increasing and decreasing bias voltage 
at a fixed ratio of $\Gamma/U=0.05$ for the symmetric Anderson model ($\varepsilon_{d}/U=-0.5$).
In equilibrium, at zero bias, 
the impurity possesses a finite magnetic moment for
the chosen parameter set, while in case of high bias voltages ($\mu > 0.5$) the stable solution
is paramagnetic. Between the two limiting cases 
a region appears, $\mu_{c1}<\mu<\mu_{c2}$, where both the magnetic and the non-magnetic solutions  
are stable. 
{\bbf We shall refer to this region as a coexistence region. }
The existence of multiple stable solutions for the occupation numbers 
in this  region is reflected in the hysteresis of the magnetization too.
The sharp decay of the magnetization shown in the inset of
Fig.~\ref{FigMuGamma} 
and the existence of a hysteresis between the critical fields
indicate  clearly a first order transition,  
predicted incorrectly  by  the MF solution.

The parameter space can thus be divided into magnetic ($M$), paramagnetic ($P$) and
coexistence ($C$) regions.
In the paramagnetic regions
a single stable solution exists only
($\langle n_{\uparrow}\rangle=\langle n_{\downarrow}\rangle$),
while in the magnetic one two stable magnetic (corresponding to 
magnetizations $\pm m$)
and an unstable paramagnetic solution can be found.
In the coexistence region two magnetic and one paramagnetic stable solutions 
and two unstable magnetic solutions exist.
Therefore, these regions can be distinguished
by the number of the solutions of Eq.~(\ref{EqMain0}). 
In the symmetric case when
$\langle n_{\uparrow}\rangle+\langle n_{\downarrow}\rangle=1$, the easiest way to
construct a "phase diagram" is to sweep possible values of
$\langle n_{\uparrow}\rangle - \langle n_{\downarrow}\rangle$ pairs, substitute them
into Eq.~(\ref{EqMain0}) and count the number of solutions
for different parameter sets.

In the paramagnetic region one can exploit the fact
that Eq.~(\ref{EqMain0})
has a non-magnetic solution for every possible parameter set.
Therefore one can substitute $\langle n_{\uparrow}\rangle=\langle n_{\downarrow}\rangle$
into Eq.~(\ref{EqMain0}) {\bbf and search only for} non-magnetic states.
The region of stability for this solution can then be determined 
through a linear stability analysis\cite{komnik,diploma}.

{\gbf 
Fig.~\ref{FigMuGamma} shows a typical
a magnetic "phase diagram"  as a function of $\mu/U$ and $\Gamma/U$ 
for the symmetric non-equilibrium AIM at $T=0$. At the ``upper critical line''
 $\mu_{c2}$ the magnetic solution becomes unstable. 
In the magnetic case, the effective levels corresponding to different spins 
are not equally occupied and lie at different energies. The magnetic solution 
becomes unstable  when the 
value of the bias voltage reaches approximately the effective energy of 
one of the two differently occupied levels. 

The critical line $\mu_{c1}$ in Fig~\ref{FigMuGamma}  marks, on the other
hand,  the border of  stable paramagnetic solutions. 
In the special case, $\epsilon_d = -U/2$ the two spin occupations are 
 $\langle n_{\uparrow,\downarrow}\rangle=0.5$ in the whole paramagnetic region, 
and the magnetic boundary equation  simplifies to
$\mu^{2}+16\Gamma^{2}-8U\Gamma/\pi=0$. This analytical result is nicely
reproduced by our  numerical stability analysis (see Fig.~\ref{FigMuGamma}).

}

\begin{figure}
\includegraphics[width=200pt]{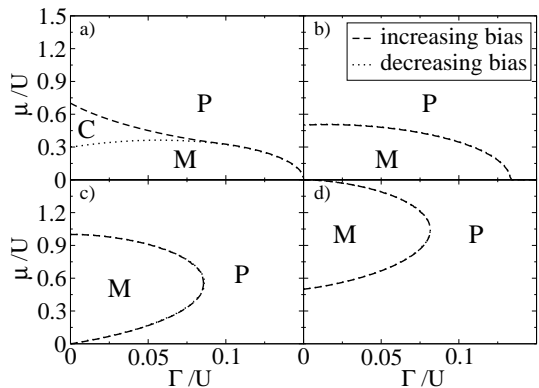}
\caption{"Phase diagrams" as a function of $\mu/U$ and $\Gamma/U$ for $T=0$ and different $\varepsilon_{d}$ values, 
$a)$ $\varepsilon_{d}/U=-0.35$ 
$b)$ 
$-0.25$, 
$c)$ 
$0$ and 
$d)$
$0.25$. 
Notation of the regions and the lines are the same as Fig.~\ref{FigMuGamma}}\label{FigDife}
\end{figure}

The values of $\mu_{c1}$ and $\mu_{c2}$ 
are functions of $\Gamma/U$ and $\varepsilon_d/U$; 
for increasing $\Gamma/U$, the coexistence and magnetic regions
disappear and only the non-magnetic solution survives. 
In Fig.~\ref{FigDife} the "phase diagrams" can be seen as a function of $\mu/U$
and $\Gamma/U$ at $T=0$.  Depending on the value of $\varepsilon_{d}$ 
we can distinguish four different regions: 
\textit{empty regime} ($\varepsilon_{d}>0$), \textit{mixed valence regime} ($\varepsilon_{d}\approx0$), 
\textit{local moment regime} ($-U\leq\varepsilon_{d}\leq0$) and a \textit{doubly occupied regime} 
($\varepsilon_{d}<-U$) which 
behaves similarly to the \textit{empty regime} by particle-hole symmetry. 

In the \textit{local moment regime}, shown in Fig.~\ref{FigMuGamma} and Fig.~\ref{FigDife}$a$,
for $\varepsilon_{d}\approx-U/2$, there is 
approximately one electron on the impurity forming a local spin moment.
In equilibrium, this finite magnetic moment is predicted on the dot
below a critical value of $\Gamma/U$, and this moment is destroyed by a large 
enough bias voltage.
The coexistence region only appears in this \textit{local moment regime} and 
vanishes above $\varepsilon_{d}\approx-U/4$ (see Fig.~\ref{FigDife}$b$).

In the \textit{empty regime} ($\varepsilon_{d}\geq0$), 
shown in Figs.~\ref{FigDife}$c$ and \ref{FigDife}$d$,
the equilibrium magnetization completely disappears. 
However, surprisingly, the MF solution predicts the appearance 
of local moments for small $\Gamma$'s and $2\varepsilon_{d}\leq\mu\leq2\varepsilon_{d}+U$ biases. 
This intriguing local moment formation has a simple physical meaning:
For large values of $U/\Gamma$ and $2\varepsilon_{d}\leq\mu\leq2\varepsilon_{d}+U$, the bias voltages 
are large enough to inject an electron to the empty level, however, 
they are not large enough to overcome the Coulomb energy of injecting a second 
electron to the local level. Therefore, electrons pass through the dot one by one, 
and a fluctuating magnetic moment appears on the dot. Note that in this regime, 
the magnetization is induced exclusively by the finite bias voltage. 

\begin{figure}
\includegraphics[width=200pt]{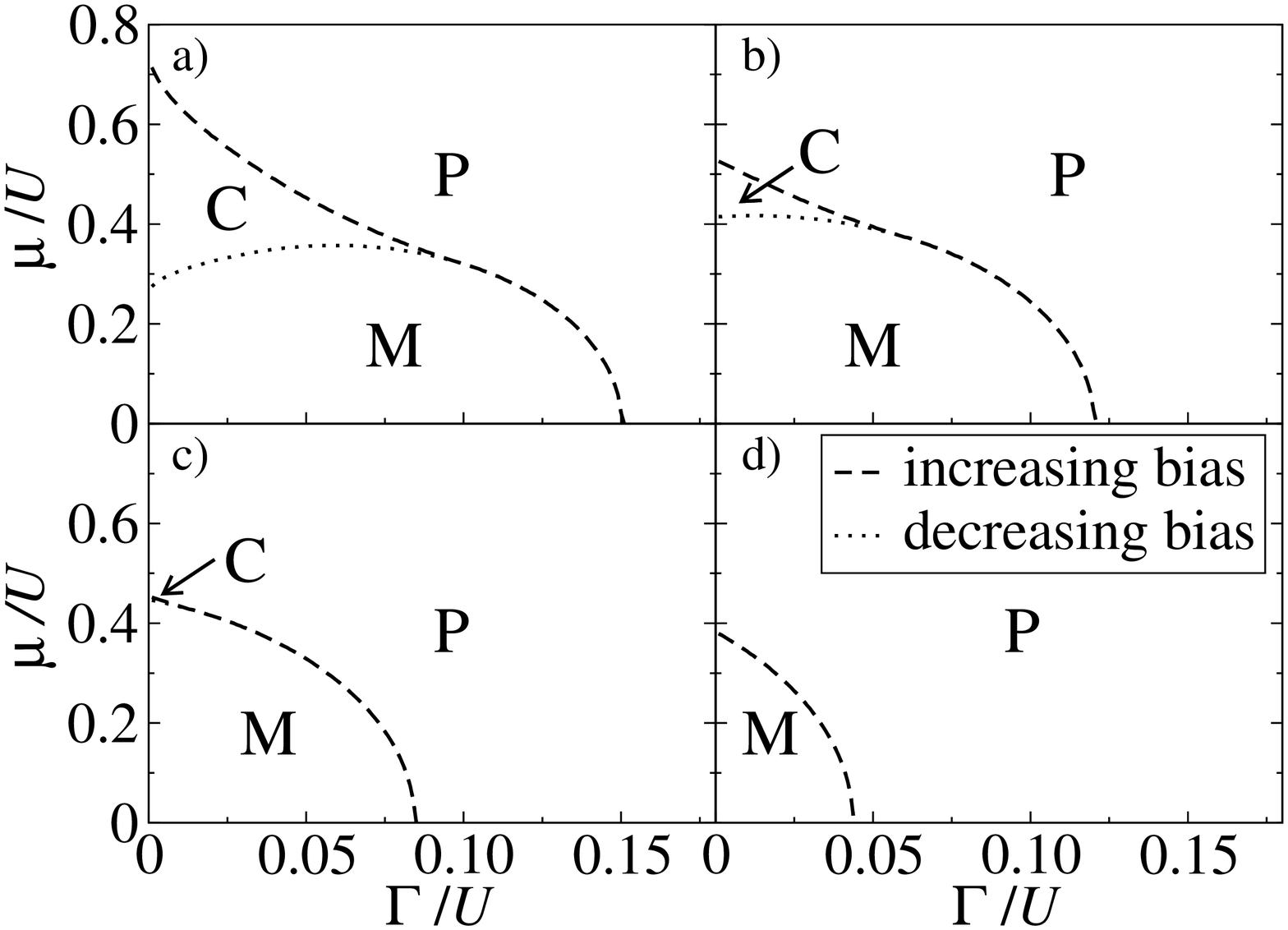}
\caption{"Phase diagrams" as a function of $\mu/U$ and $\Gamma/U$ for $\varepsilon_{d}/U=-0.5$. 
The temperature 
$a)$, 
$T/U=0.051$ 
$b)$, 
$0.101$ 
$c)$ 
$0.151$ and
$d)$
$0.201$.
Notation of the regions and the lines are the same as
Fig.~\ref{FigMuGamma}}\label{FigTemp}
\end{figure}

The overall effect of the temperature in the applied MF-approximation 
is to destroy the magnetic moment on the impurity 
and drive the system to be paramagnetic.
Fig.~\ref{FigTemp} shows the temperature dependence of the "phase diagram"
in the symmetric case. The coexistence region gradually vanishes for increasing temperatures, 
while the magnetic and paramagnetic regions get larger.
Increasing the temperatures further the magnetic region disappears too.

\begin{figure}
\includegraphics[width=200pt]{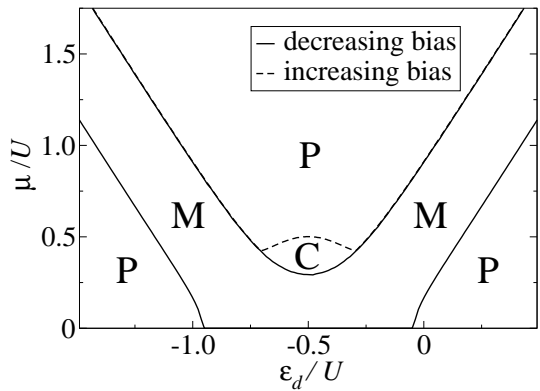}
\caption{
Magnetic (M), coexistence (C) and paramagnetic (P) regions
as a function of $\mu/U$ and $\varepsilon_{d}/U$ values, for $\Gamma/U=0.05$.
}\label{muvar}
\end{figure}

The previous results are summarized  in Fig.~\ref{muvar} 
for a fixed ratio $\Gamma/U=0.05$ that is
already sufficiently small to find the coexistence and magnetic regions
in the local moment regime.
The whole "phase diagram" is symmetric to $\varepsilon_{d}/U=-0.5$
due to the electron-hole symmetry. 
Note that the coexistence region only exists close to the electron-hole 
symmetry, $\varepsilon_{d}/U\approx-0.5$.
For large asymmetries (double or zero equilibrium occupation)
a non-equilibrium magnetic solution
appears, while in equilibrium only the paramagnetic solution exists. 
This magnetic region has also been observed although not discussed 
in detail in Ref. \onlinecite{komnik}.

The mean-field solution also leads to the appearance of non-physical 
features in the transport properties of the impurity. 
We calculated the transport properties within the MF approximation 
using the Landauer-B\"uttiker formalism\cite{landauer}, but 
similar results can be obtained applying the Keldysh formalism\cite{diploma}.
The current can be evaluated as
\begin{equation}\label{current}
I_{\sigma}\equiv\frac{e}{h}\int_{-\infty}^{\infty}
\mathrm{d}\varepsilon\left(f_{L}(\varepsilon)-f_{R}
(\varepsilon)\right)|t_{\sigma}(\varepsilon)|^{2}\;.
\end{equation}
Here the transmission coefficient  
$t_{\sigma}= \beta_{\sigma}
\sqrt{v_{k'}/v_{k}}$ 
is normalized to the flux, 
with  $v_{k}$ and $v_{k'}$  the velocities of incident and transmitted 
electrons with wave numbers $k$ and $k'$.
Applying the large bandwidth approximation again,
the current can be written for finite temperatures as
\begin{equation}
I_{\sigma}=
\frac{e}{2\pi\hbar}\int\limits_{-\infty}^{\infty}\mathrm{d}\varepsilon
\frac{\Gamma^{2}\left(f_{L}(\varepsilon)-f_{R}(\varepsilon)\right)}
{\left(\varepsilon-\varepsilon_{d}-U\langle n_{-\sigma}
\rangle\right)^{2}+\Gamma^{2}}\label{EqCur2}\;,
\end{equation}
with $\langle n_{\sigma}\rangle$ the non-equilibrium occupation numbers obtained from Eq.~(\ref{EqMain}). 
This integral can be trivially evaluated at $T=0$ temperature.

\begin{figure}[t]
\includegraphics[width=200pt]{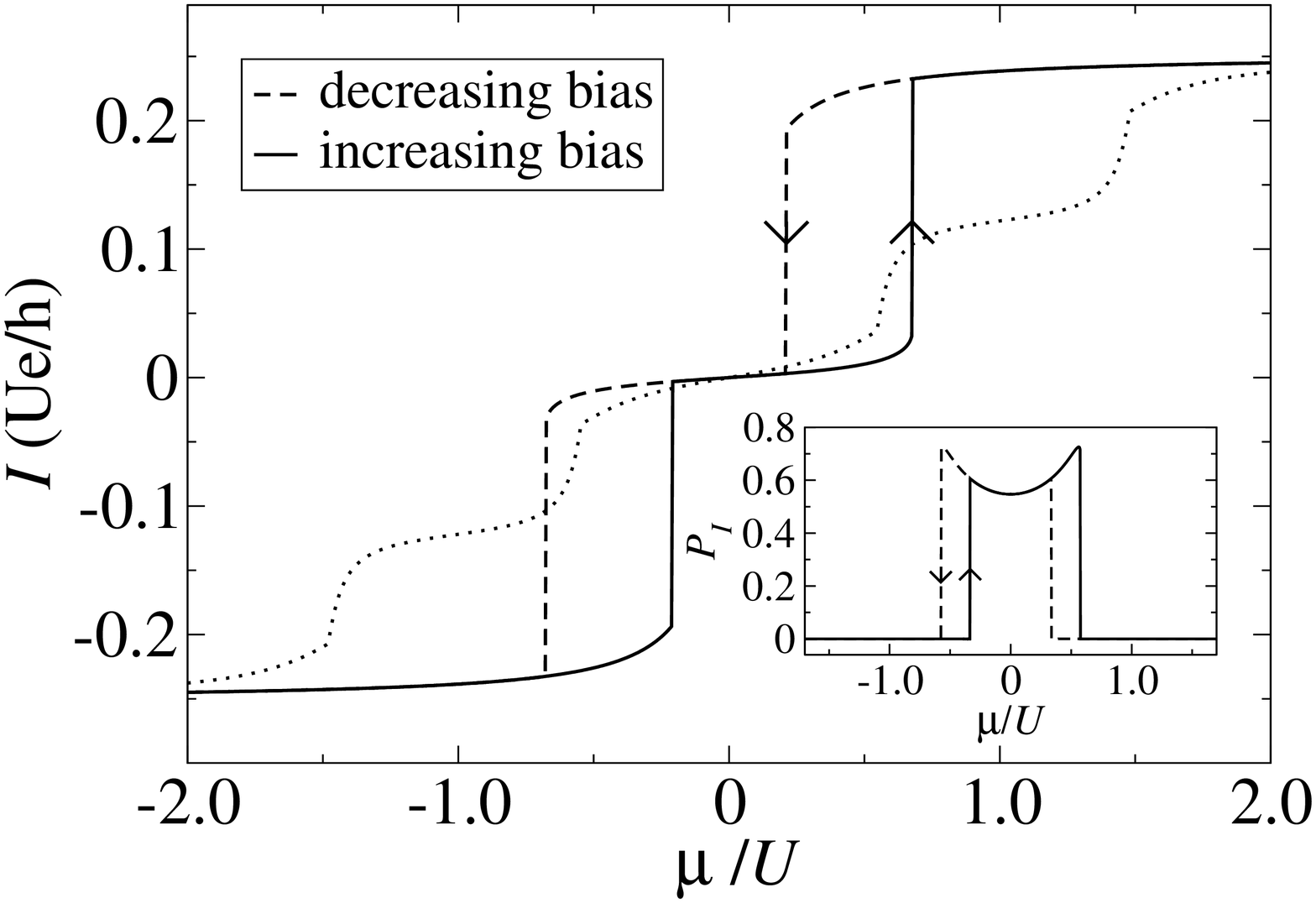}
\caption{
\label{FigCur}
The current as a function of the bias voltage for
$\varepsilon_{d}/U=-0.5$ (solid and dashed line) 
and $\varepsilon_{d}/U=0.25$ (dotted line)
for $\Gamma/U=0.02$.  Hysteresis can be observed in the current in the \textit{local moment regime}. 
Inset: the polarization of 
the current for $\varepsilon_{d}/U=-0.35$ and $\Gamma/U=0.02$.}
\end{figure}

Fig.~\ref{FigCur} shows the current as a function of bias 
for two different level positions in the strongly correlated regime.
For $\varepsilon_{d}/U=-0.5$ we find  hysteresis 
in the current. For $\varepsilon_{d}/U=0.25$, on the other hand, 
no hysteresis appears but  the current shows a two-step behavior as a function of
bias voltage. In this \textit{empty regime}, the MF 
equations thus account qualitatively correctly for the charging of the 
local level, but the kinks appearing in the $I(\mu)$ curve are due to the 
incorrectly predicted symmetry breaking and are thus again 
artifacts of the MF solution.

In the inset of Fig.~\ref{FigCur} we have plotted the polarization of the current,
$P_{I}\equiv (I_{\uparrow}-I_{\downarrow})/(I_{\uparrow}+I_{\downarrow})$, 
as a function of $\mu$. 
In the symmetric case the current is not polarized  due to the electron-hole
symmetry, but  for small electron-hole asymmetries, a hysteresis appears in
the polarization too.  In general,  the polarization is finite whenever a local moment 
appears on the impurity and there is no electron-hole symmetry. 

To conclude, we have studied the Anderson model out of equilibrium in 
the framework of the scattering formalism combined with a mean-field approximation. 
This method, frequently used in molecular transport calculations, 
 incorrectly predicts a magnetic phase transition  as well as a bias-induced 
magnetic moment formation, accompanied by 
hysteresis in various physical quantities and the coexistence of 
multiple solutions. The MF approach thus fails 
whenever correlations become important. These artifacts of the mean field approach 
should alert physicists who study  transport through strongly correlated 
and magnetic molecules, and urge one to use more sophisticated methods
that avoid spontaneous symmetry breaking and account for  
dynamical effects.

\textit{Acknowledgment.}
We are grateful L\'aszl\'o Szunyogh for valuable discussions and remarks. This research has been supported by Hungarian Grants
No. OTKA 
NF061726, 
T046303, 
and F68726. 

\end{document}